\documentclass[10pt,letterpaper,twoside,twocolumn,journal,letterpaper, final,twosided]{IEEEtran}
\usepackage[T1]{fontenc}
\usepackage[utf8]{inputenc}
\usepackage{color}
\usepackage{prettyref}

\usepackage{bm}
\usepackage{amsmath}
\usepackage{amssymb}
\usepackage{graphicx}
\usepackage{xfrac}

\makeatletter

\pdfpageheight\paperheight
\pdfpagewidth\paperwidth

\@ifundefined{date}{}{\date{}}
\usepackage{graphicx}
\setkeys{Gin}{width=\linewidth}
\usepackage{slashbox}
\newcommand{\berkerstitle}{Non-Redundant~OFDM~Receiver~Windowing for~5G~Frames~\&~Beyond}
\usepackage{color} 
\definecolor{mygreen}{RGB}{28,172,0} 
\definecolor{mylilas}{RGB}{170,55,241}

\ifCLASSOPTIONcompsoc
\usepackage[caption=false,font=normalsize,labelfont=sf,textfont=sf]{subfig}
\else
\usepackage[caption=false,font=footnotesize]{subfig}
\fi

\usepackage{mathtools}
\usepackage{newtxmath}

\newcommand{\toep}[2]{\ensuremath{\operatorname{\mbox{T\hspace{-.5em}-}}\left({#1},{#2}\right)}}
\newcommand{\kron}[2]{\ensuremath{{#1} \otimes {#2}}}
\newcommand{\diag}[1]{\operatorname{diag}\left({#1}\right)}

\newcommand*{\matr}[1]{\mathbf{#1}}
\newcommand*{\matrgr}[1]{\bm{#1}}
\newcommand*{\tran}[1]{{#1}^{\mkern-1.5mu\mathsf{T}}}
\newcommand*{\conj}[1]{{#1}^{\mkern-1.5mu \ast}}
\newcommand*{\herm}[1]{{#1}^{\mkern-1.5mu\mathsf{H}}}
\newcommand*{\osq}[1]{\ensuremath{{#1}\odot\conj{#1}}}

\usepackage{dblfloatfix}
\usepackage{balance}
\usepackage[caption=false,font=footnotesize]{subfig}
\usepackage{cite}

\usepackage{prettyref}
\newrefformat{fig}{Fig.~\ref{#1}}
\newrefformat{conj}{Conjecture~\ref{#1}}
\newrefformat{tab}{Table~\ref{#1}}
\newrefformat{sec}{Section~\ref{#1}}
\newrefformat{subsec}{Section~\ref{#1}}
\newrefformat{subsubsec}{Section~\ref{#1}}

\newcounter{tempEquationCounter} 
\newcounter{thisEquationNumber}
\newenvironment{floatEq}
{\setcounter{thisEquationNumber}{\value{equation}}\addtocounter{equation}{1}
\begin{figure*}[!t]
\normalsize\setcounter{tempEquationCounter}{\value{equation}}
\setcounter{equation}{\value{thisEquationNumber}}
}
{\setcounter{equation}{\value{tempEquationCounter}}
\hrulefill\vspace*{4pt}
\end{figure*}
}

\usepackage{acronym}
\input{acros}
\usepackage{textcomp}
\usepackage{algorithmic}

\usepackage{siunitx}
\DeclareSIUnit{\belmilliwatt}{Bm}
\DeclareSIUnit{\dBm}{\deci\belmilliwatt}
\DeclareSIQualifier{\isotropic}{i}
\DeclareSIQualifier{\carrier}{c}

\makeatother

\usepackage{mathpazo}
\usepackage[mathpazo]{flexisym}
\usepackage{breqn}
\setkeys{breqn}{labelprefix={eq:}}

\begin{document}
\bstctlcite{BSTcontrol}
\title{\berkerstitle}

\author{Berker~Peköz\IEEEmembership{,~Graduate Student Member, IEEE,} Z.~Esat~Ankaralı,
	 Selçuk~Köse\IEEEmembership{,~Member,~IEEE}
	and Hüseyin~Arslan\IEEEmembership{,~Fellow,~IEEE}
	\thanks{Copyright \copyright 2015 IEEE. Personal use of this material is permitted. However, permission to use this material for any other purposes must be obtained from the IEEE by sending a request to pubs-permissions@ieee.org.}
	\thanks{This work was supported in part by the National Science Foundation under Grant 1609581.}
\thanks{Berker~Peköz is with the Department of Electrical Engineering, University of South Florida, Tampa, FL 33620 USA. (e-mail: pekoz@usf.edu)}
	\thanks{Zekeriyya~Esat~Ankaralı was with the Department of Electrical Engineering, University of South Florida, Tampa, FL 33620 USA. He is now with Maxlinear Inc., Carlsbad, CA 92008 USA. (e-mail: zekeriyya@mail.usf.edu)}
	\thanks{Selçuk~Köse is with the Department of Electrical and Computer Engineering, University of Rochester, Rochester, NY 14627 USA. (e-mail: selcuk.kose@rochester.edu)}
	\thanks{Hüseyin~Arslan is with the Department of Electrical Engineering, University of South Florida, Tampa, FL 33620 USA and also with the Department of Electrical and Electronics Engineering, Istanbul Medipol University, Istanbul, 34810 TURKEY. (e-mail: arslan@usf.edu)}}
\maketitle
\begin{abstract}
Contemporary \ac{rw-ofdm} algorithms have limited \ac{aci} rejection capability under high delay spread and small \ac{fft} sizes. \Ac{cp} is designed to be longer than the \ac{med} of the channel to accommodate such algorithms in current standards. The robustness of these algorithms can only be improved against these conditions by adopting additional extensions in a new backward incompatible standard. Such extensions would deteriorate the performance of high mobility vehicular communication systems in particular. In this paper, we present a low-complexity Hann \ac{rw-ofdm} scheme that provides resistance against \ac{aci} without requiring any \ac{isi}-free redundancies. While this scheme is backward compatible with current and legacy standards and requires no changes to the conventionally transmitted signals, it also paves the way towards future spectrotemporally localized and efficient schemes suitable for higher mobility vehicular communications. A Hann window effectively rejects unstructured \ac{aci} at the expense of structured and limited \ac{ici} across data carriers. A simple \ac{mrc}-\ac{sic} receiver is therefore proposed to resolve this induced \ac{ici} and receive symbols transmitted by standard transmitters currently in use. The computational complexity of the proposed scheme is comparable to that of contemporary \ac{rw-ofdm} algorithms, while \ac{aci} rejection and \ac{ber} performance is superior in both long and short delay spreads. Channel estimation using Hann \ac{rw-ofdm} symbols is also discussed.\acresetall
\end{abstract}

\begin{IEEEkeywords}
5G mobile communication, interference cancellation, interference elimination, multiple access interference, numerology
\end{IEEEkeywords}

\section{Introduction\label{sec:introduction}}
Next generation cellular communication standards beyond 5G mobile communication are planned
to schedule non-orthogonal sub-frames, referred to as numerologies,
in adjacent bands\cite{ankarali2017flexible}. Numerologies, in their current definition, refer to \ac{cp-ofdm} waveforms using different subcarrier spacings, and in some cases, various \ac{cp} rates\cite{3gpp.38.211}. Different numerologies
interfere with one-another\cite{Zhang2018} and \ac{aci} becomes the factor limiting data rates if the interfering block outpowers the desired block at the intended receiver\cite{lupas1990nearfar}. 

Nodes can reject \ac{aci} by filtering\cite{abdoli_filtered_2015}
or windowing\cite{muschallik1996improving} the received signal. Filtering
requires matched filtering operation at the both ends of the communication system for optimal
performance\cite{turin_introduction_1960}. If not already implemented at both nodes,
this requires modifying the device lacking this function, which is unfeasible
for \acp{ue} that are produced and in-use. The additional complex
multiplication and addition operations required to filter the signal
increase the design complexity of the modem, which in
turn increase the chip area, production cost, power consumption, and
operational chip temperature and reduces the lifetime of the device
and battery\cite{vaisband2016onchippower}. Introducing these operations at the \ac{gnb} can be justified to improve system performance, however the takeaways may cause \ac{iot} devices to fall short of their \acp{kpi} and is undesirable \cite{Gozalvez2016}.

\begin{figure}
	\subfloat[]{
		\includegraphics{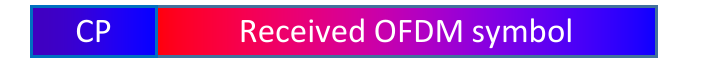}
		\label{fig:stdsym}}\hfil
	\subfloat[]{
	\includegraphics{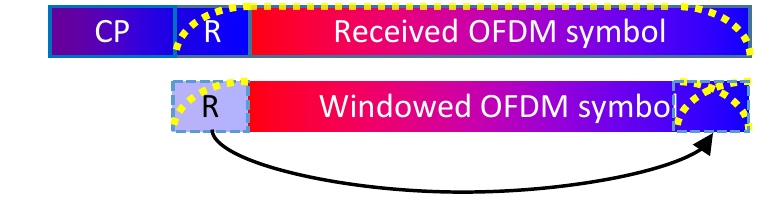}
	\label{fig:convwinsym}}\hfil
	\subfloat[]{
	\includegraphics{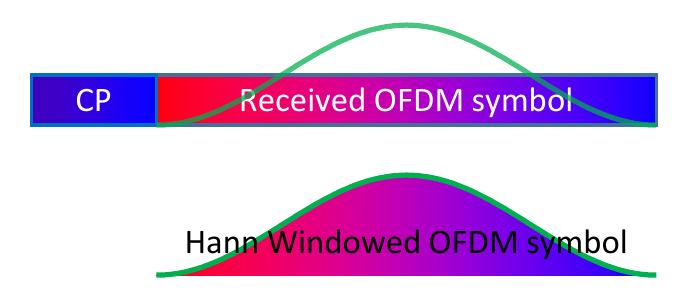}
	\label{fig:hannsym}}\hfil
	\caption{(a) Standard symbol structure, (b) Symbol structure needed to utilize conventional receiver windowing algorithms effectively in channels with long delay spread, and (c) Hann receiver windowing using standard symbol structures.}
	\label{fig:symstr}
\end{figure}

Receiver windowing is another method proposed to reduce \ac{aci} absorption\cite{muschallik1996improving} and is extensively studied in the literature\cite{bala_shaping_2013}.
Conventional \ac{rw-ofdm} algorithms require an abundant periodic extension of the transmitted signal that is free from multipath echoes of the previous symbol to maintain orthogonality of the system\cite{guvenkaya2015awindowing}. However, such an extension may not always be available or may be little, especially in vehicular communication channels requiring shorter symbol durations\cite{Mecklenbrauker2011}. 
Trying to utilize these algorithms in these conditions would require adding an additional extension, as shown in \prettyref{fig:convwinsym}.
However, modifying the symbol structure defined in both 4G and 5G mobile communication standards \cite{3gpp.38.211}, shown in \prettyref{fig:stdsym}, with such an extension breaks orthogonality with all other devices that use the standard frame structure\cite{ankarali2017flexible}. 
Even if any gain for the desired user itself can be made possible by incorporating such extension for receiver windowing, introducing such elevated interference to others is not allowed by the current standards\cite{3gpp.38.912}. Furthermore, both ends of the communication must be aware of and agree to make such a change. 

Another potential problem regarding additional extensions for windowing is the increase in the effective symbol duration which reduces the effective symbol rate. Due to the time variation of the channel in high mobility systems, the additional extensions not only cause a direct reduction in data rate but also either further cuts the data rate back when relative pilot overhead is increased to mitigate the reduction in absolute pilot periodicity or reduces capacity due to the channel estimation errors when no modification is done\cite{wu_survey_2016}. In order to achieve reliable high mobility vehicular communications, there is an apparent need to shorten the cyclic extensions instead of further elongating them.

\begin{figure}
	\subfloat[]{
		\includegraphics{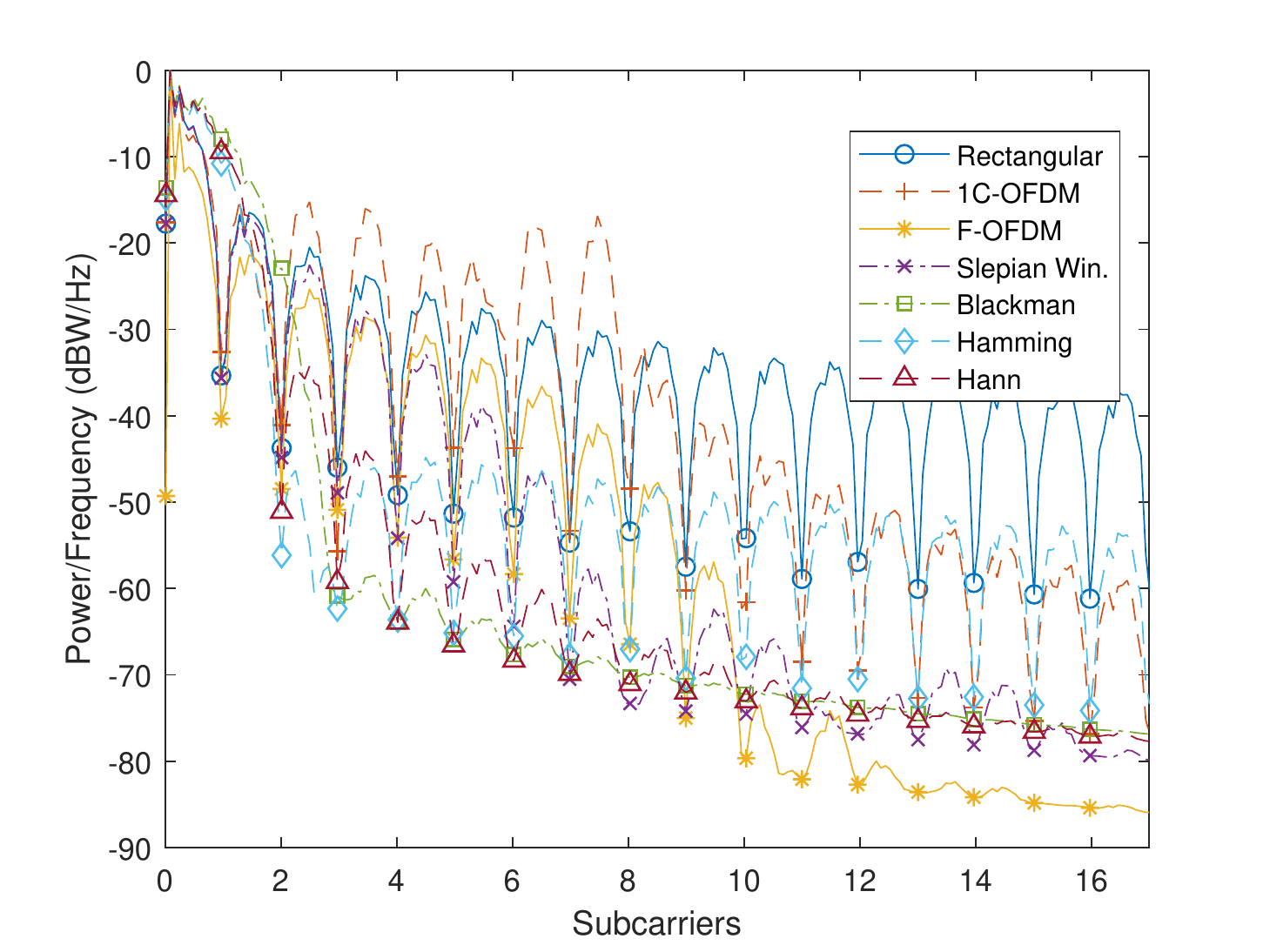}
	\label{fig:psdlongcp}}\hfil
	\subfloat[]{
		\includegraphics{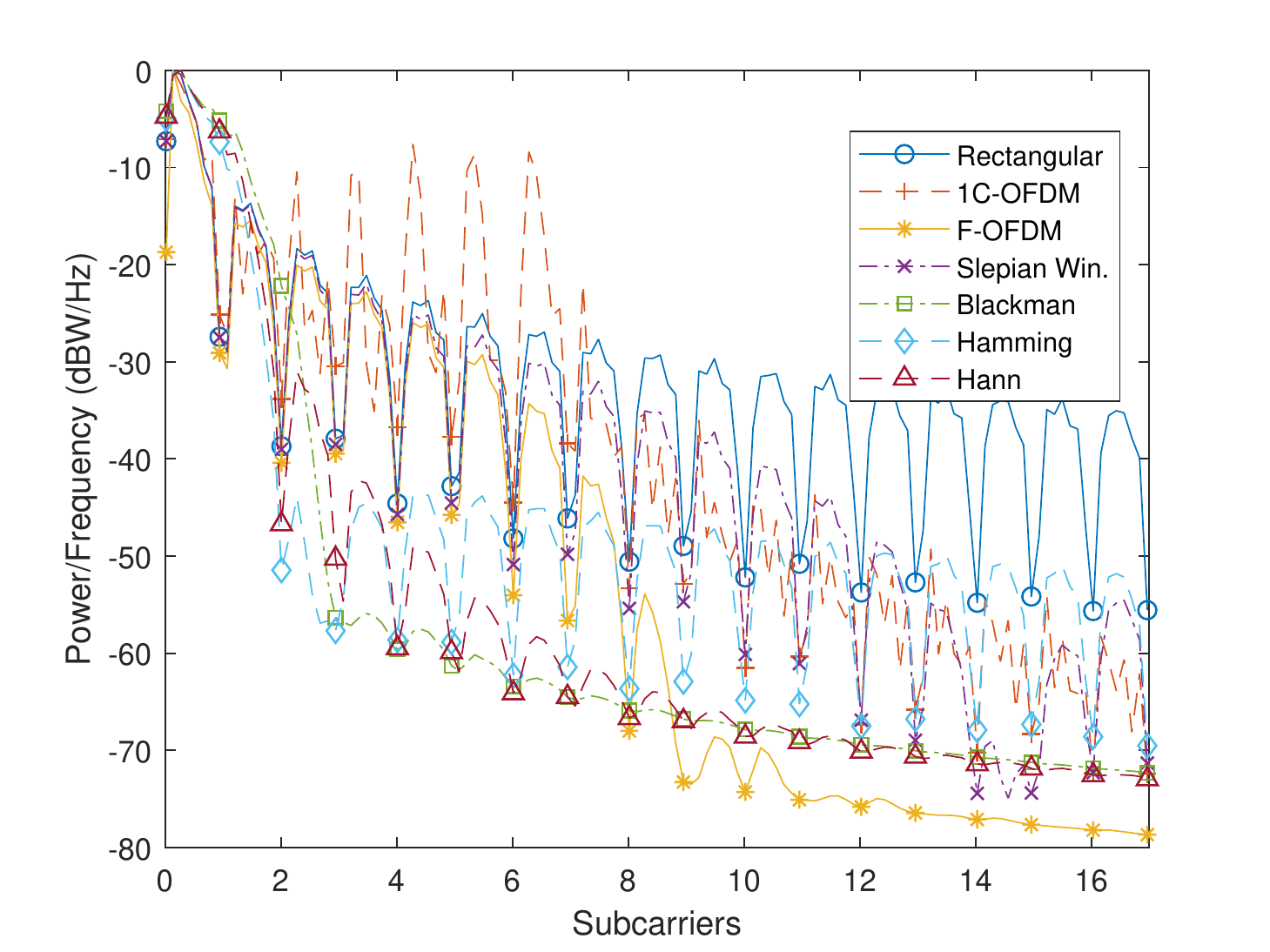}
	\label{fig:psdshortcp}}
	\caption{The \acp{psd} of \ac{ofdm} schemes and window functions applied to a (a) long duration \ac{ofdm} symbol with extended \ac{cp} and (b) short duration \ac{ofdm} symbol with normal \ac{cp} rates.
		 The markers indicate \ac{fft} sampling points.}\label{fig:psd}
\end{figure}

A receiver windowing approach that utilizes the \ac{cp} disturbed by multipath interference to reject \ac{aci} while conserving the legacy frame structure was proposed in \cite{pekoz2017adaptive}. Reducing \ac{aci} with this approach comes at the cost of introducing \ac{isi}, which consists of the sum of the low powered contributions from all subcarriers of the previous symbol. The computational complexity of canceling the \ac{isi} is high due to the large number of interfering components. Furthermore, this approach is not effective with shorter \ac{cp} durations that are associated with vehicular communication numerologies. Consider \prettyref{fig:psd}, which shows the \acp{psd} of the the sixth subcarrier from the band-edge of different \ac{ofdm} variations and window functions.
 The \ac{psd} labeled as "Slepian\cite{slepian1961prolate} Win." in \prettyref{fig:psdlongcp} is obtained by performing receiver windowing operation presented in \cite{guvenkaya2015awindowing} on an extended \ac{cp} numerology\cite{3gpp.38.211} using the entire \ac{cp} duration of a small subcarrier spacing, long duration \ac{ofdm} symbol. In this case, the window works as expected and is able to confine the spectrum within the \ac{rb} as intended, and consistently fades throughout the spectrum. However, if the same algorithm is applied to a short duration vehicular numerology with normal \ac{cp} overhead, as shown in \prettyref{fig:psdshortcp}, the window underperforms and provides a limited benefit over the standard rectangular window even if the whole \ac{cp} duration is still used. Furthermore, the \ac{psd} behaves inconsistently throughout the spectrum due to the limited resolution especially for the subcarriers of the edgemost \ac{rb} as presented, oscillating to high powers away from the subcarrier of interest.
  Also note that this is the performance upper bound for a normal \ac{cp} overhead. If a shorter window duration is used to utilize part of the \ac{cp} for its actual purpose to mitigate multipath channel and limit \ac{isi}, the performance reduces further. \Ac{fofdm} \cite{abdoli_filtered_2015} does not suffer from the same problem, but requires changes at the transmitting device and is computationally complex. \ac{ncofdm} \cite{beek2009ncontinuous} can be utilized by all devices in the band to consistently reduce the \ac{aci} levels regardless of symbol duration, but this scheme also requires changes at both transmitting and receiving devices, and also introduces in-band interference as seen in \prettyref{fig:psd}. There is an apparent need for a reception algorithm that does not modify the standard transmitter, has low computational complexity, and is robust against delay spread without requiring extensions, and is not affected by the FFT size.

In a regular \ac{ofdm} based system, if no redundancy is used for windowing, and a receiver window function other than rectangular is used, the zero crossings of the window's frequency response
differs from that of the transmitted subcarriers\cite{bala_shaping_2013}. This causes
heavy \ac{ici} between the received subcarriers, resulting in problems greater than the avoided \ac{aci} \cite{muschallik1996improving}. Attempting to cancel the resulting \ac{ici} yields little return if the \ac{ici} consists of weak contributions from numerous subcarriers, and the computational complexity and success of the cancellation renders such implementation impractical in general. Some window functions commonly used in signal processing reveal special cases\cite{blackman_measurement_1958} if the windowing operation depicted in \prettyref{fig:hannsym} is performed, limiting the number of interfering subcarriers which may be exploited to possibly enable gains. A strong candidate is the Blackman window function, which provides promising \ac{aci} rejection seen in \prettyref{fig:psd}. However, the main lobe of the Blackman window function spans 2 adjacent subcarriers on the shown right hand side spectrum and 2 more on the not-visible left hand side, thereby including high-power \ac{ici} from a total of 4 subcarriers.  This results in computationally intensive reception and limits capacity gains. Another strong candidate is the Hamming window function, which only interferes with the closest 2 adjacent subcarriers, hence enabling lower-complexity reception. The ACI rejection performance of Hamming window function in the subcarriers that immediately follow the main lobe is also unmatched. However, considering the ACI rejection performance throughout the rest of the spectrum and the power of the inflicted ICI due to windowing, the Hann window function is distinguished from other candidates and is chosen in this study  to satisfy this apparent need.
 A similar investigation during the design of the \ac{gsm} system led in favor of the \ac{gmsk} pulse shapes that are inherently non-orthogonal only with a finite number of symbols around them and \ac{sir} degradation is manageable in severe multiple access multipath channel conditions, instead of other candidates that are ideally orthogonal but suffer severe \ac{sir} degradation once this orthogonality is lost due to multiple access multipath channel \cite{Murota1981}.
Because of the aforementioned spectral features, Hann windowing similarly converts a complex \ac{aci} problem, with its out-of-band rejection performance comparable to optimum windowing as shown in \prettyref{fig:psdlongcp}, to a manageable \ac{ici}  problem requiring little computational complexity at the receiver\cite{harris1978onthe,honig2009overview}.

\begin{figure*}
	\includegraphics{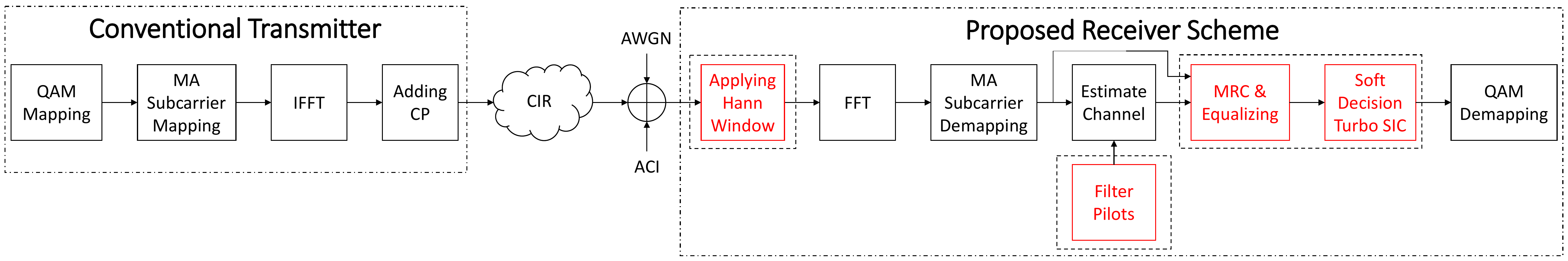}
	\caption{The block diagram of the proposed scheme, highlighting the modifications to the standard receiver structure with dashed blocks (standard transmitter is not modified).}\label{fig:blockdiag}
\end{figure*}

We present a novel transceiver algorithm that mitigates the \ac{ici} resulting from Hann windowing. This algorithm performs well regardless of \ac{ofdm} symbol duration, \ac{cp} duration or delay spread. The algorithm is solely a receiver algorithm that can be used to receive the signals transmitted from a conventional legacy transmitter using any modulation. Therefore the systems using either of the proposed algorithms are interoperable with future and legacy standards. This algorithm consists of maximizing \ac{sinr} first using \ac{mrc}, afterwards mitigating the \ac{ici} using a soft decision turbo \ac{sic} equalizer. Furthermore, the computational complexity of the algorithm is less than or comparable to \cite{guvenkaya2015awindowing}, while a higher performance is achieved in most conditions. A block diagram of the proposed method is presented in \prettyref{fig:blockdiag}.

 Our contributions in this work are as follows:
\begin{itemize}
	\item A redundancy free \ac{rw-ofdm} scheme that outperforms prior art without requiring changes to the standard frame structure regardless of channel conditions is proposed. The proposed scheme has high \ac{aci} rejection performance at the expense of a structured \ac{ici} that can be resolved without computationally intensive computations.
	\item A channel estimation technique of Hann \ac{rw-ofdm} symbols and 5G mobile communication system pilots is proposed.
	\item The \ac{ici} contribution from and to each subcarrier resulting from application of a Hann window to a received \ac{ofdm} signal is derived.
	\item \ac{mrc} coefficients maximizing the \ac{sinr} of a Hann \ac{rw-ofdm} receiver as a function of the \ac{aci}, noise power and \ac{ici} is derived.
	\item The \ac{ici} contribution from and to each subcarrier resulting from application of \ac{mrc} is derived.
	\item The computational complexity of the proposed scheme is derived.
\end{itemize}

The rest of this article is organized as follows: The system model
is provided in \prettyref{sec:System-Model}, the proposed methods
are detailed in \prettyref{sec:Proposed-Method}, the interference reduction
and capacity improvement characteristics of Hann \ac{rw-ofdm} are presented in \prettyref{sec:Numerical-Results}.
The paper is concluded in \prettyref{sec:Conclusion}. 

\textit{Notation}: $\tran{\left(\cdot\right)}$, $\conj{\left(\cdot\right)}$
and $\herm{\left(\cdot\right)}$ denote the transpose, complex conjugate,
and Hermitian operations, $ \matr{e}_{i,N} $ corresponds to the $ i $th row of the $N\times N$ identity matrix $ \matr{I}_{N} $, $\matr{A}\odot\matr{B}$ and $\matr{A}\oslash\matr{B}$
correspond to Hadamard multiplication and division of matrices $\matr{A}$
and $\matr{B}$, and $\matr{A}$ by $\matr{B}$, $\matr{0}_{a\times b}$
and $\matr{1}_{a\times b}$ denote matrices of zeros and ones with
$a$ rows and $b$ columns, $\diag{\matr{v}}$ returns
a square diagonal matrix with the elements of vector $\matr{v}$
on the main diagonal, $\diag{\matr{M}}$ returns
the elements on the main diagonal of matrix $\matr{M}$ in a vector, $\mathcal{CN}\left(\mu,\sigma^{2}\right)$
represents complex Gaussian random vectors with mean $\mu$ and variance
$\sigma^{2}$, $\toep{\matr{c}}{\matr{r}}$ yields the
Toeplitz matrix where the first column is $\matr{c}$ and the first row
is $\matr{r}$, $\kron{\matr{A}}{\matr{B}}$ is the Kronecker
tensor product of $\matr{A}$ and $\matr{B}$ matrices.

\section{System Model\label{sec:System-Model}}
We aim to receive the information transmitted by a user,
hereinafter referred to as the desired user, of which corresponding elements are distinguished with subscript 0 in the multi-user context. The desired user is transmitting
data over $D$ contiguous subcarriers in an $N$ subcarrier \ac{cp-ofdm} system.
To prevent \ac{isi} across consecutive \ac{ofdm} symbols and to
transform the linear convolution of the multipath channel to a circular
convolution, a \ac{cp} of length $L$ samples is prepended to each transmitted \ac{ofdm} symbol. The samples corresponding to a \ac{cp-ofdm}
symbol of the desired user are denoted by $\matr{x}_0\in\mathbb{C}^{\left(N+L\right)\times1}$, and
are obtained as $\matr{x}_0=\matr{A}\herm{\matr{F}_{N}}\matr{M}\matr{d}$,
where $\matr{F}_{N}\in\mathbb{C}^{N\times N}$ is the $N$-point
\ac{fft} matrix, $\matr{M}\in\mathbb{Z}^{N\times D}$ is the subcarrier
mapping matrix, $\matr{d}\in\mathbb{C}^{D\times1}$ is the \ac{sc} modulated data vector
to be transmitted and $\matr{A}\in\mathbb{R}^{\left(N+L\right)\times N}$ is
the \ac{cp} addition matrix defined as
\begin{equation}
\matr{A}=\begin{bmatrix}\begin{array}{cc}
\matr{0}_{L\times\left(N-L\right)} & \matr{I}_{L}\end{array}\\
\matr{I}_{N}
\end{bmatrix}.
\end{equation}

During the transmission of the desired user, the adjacent bands are employed for communication by other users, hereinafter referred to as interfering users, of which signaling is neither synchronous nor orthogonal to that of the desired user. The signals transmitted from all users propagate through a time varying multipath channel before  reaching the receiver. Assuming perfect synchronization to the desired user's signal, let the channel gain of the $ k $th sample of the desired and $ j $th interfering user's signals, for $  j \neq0 $, during the reception of the $ n $th sample be denoted by $ h_{0,n,k} $ and $ h_{ j ,n,k} $, respectively.
 For clarity, we assume that $ \sum_{k=1}^{N+L}\lvert h_{ j ,n,k}\rvert^2=1,\,\forall j  $. If the channel convolution matrix of the $ j $th user for the scope of the desired user's symbol of interest is shown with  $\matr{H}_ j \in\mathbb{C}^{\left(N+L\right)\times\left(N+L\right)}$, respectively; the element in the $k$th column of $n$th row of any $\matr{H}_ j $ is $ h_{ j ,n,k} $, respectively. It should be noted that, if $  j  $th user's channel was time-invariant, $\matr{H}_ j $ would be a Toeplitz matrix, whereas in this model, the elements are varying for all $  j  $ and the autocorrelation functions and the power spectral densities of any diagonal of any channel convolution matrix fit those defined in \cite{gans1972apowerspectral}.
The first $N+L$ samples received over the wireless medium under perfect synchronization to the desired user's signal normalized to the noise power are stored in $\matr{y}\in\mathbb{C}^{\left(N+L\right)\times1}$, which is given as
\begin{equation} \label{eq:recsamps}
\matr{y}= z+\sum_{ j }\left(\sqrt{\gamma_{ j }}\matr{H}_{ j }\matr{x}_{ j }\right), 
\end{equation}
 where 
 $z\sim\mathcal{CN}\left(0,1\right)$ is the background \ac{awgn},
 $ \gamma_0 $ and $ \gamma_ j  $ are the \acp{snr} of the desired and $ j $th interfering user, respectively, and
 $\matr{x}_{ j }$ is the sample sequence transmitted by $ j $th interfering user in the reference duration of the desired symbol for $  j \neq0 $.
 
 \subsection{Reception in Self-Orthogonal \ac{rw-ofdm} Systems}
A brief review of channel estimation, equalization and \ac{ici} in self-orthogonal conventional \ac{rw-ofdm} systems may help understand the derivation of the aforementioned for Hann \ac{rw-ofdm}.

For the sake of brevity, assume the receiver utilizes an extensionless receiver windowing function of tail length $K\in\mathbb{N}_{\leq L}$ for all subcarriers to receive the data transmitted by the desired user, and the window function coefficients scaling the \ac{cp} are shown as $ \matr{\breve{w}}_K\in\mathbb{R}^{K\times1} $\cite{pekoz2017adaptive}. Then, the windowed \ac{cp} removal matrix $ \matr{B}_K\in\mathbb{R}^{N\times\left(N+L\right)} $ is obtained as shown in \prettyref{eq:winMat}. Note that for $ K=0 $, \prettyref{eq:winMat} reduces to the rectangular windowing \ac{cp} removal matrix $ \matr{B}_0=\begin{bmatrix}
\matr{0}_{N\times L} & \matr{I}_N
\end{bmatrix} $.
\begin{floatEq}
	\begin{equation}
		\matr{B}_K=\begin{bmatrix}
		\matr{0}_{\left(N-K\right)\times \left(L-K\right)} & \matr{0}_{\left(N-K\right)\times K} & \matr{I}_{\left(N-K\right)\times \left(N-K\right)} & \matr{0}_{\left(N-K\right)\times K}\\
		\matr{0}_{K\times \left(L-K\right)} & \diag{\matr{\breve{w}}_K} & \matr{0}_{K\times \left(N-K\right)} & \matr{I}_K-\diag{\matr{\breve{w}}_K}
		\end{bmatrix} \label{eq:winMat}
	\end{equation}
\end{floatEq}
The received symbols in a \ac{rw-ofdm} system are then given as
\begin{align}
	\matr{r}&=\tran{\matr{M}}\matr{F}_N\matr{B}_K\matr{y}\\
	&=\tran{\matr{M}}\matrgr{\Theta}\matr{d}+\tran{\matr{M}}\matr{\tilde{z}},
\end{align}
where the channel disturbance vector $\matr{\tilde{z}}\in\mathbb{C}^{N\times1}$ is
\begin{align}
	\matr{\tilde{z}}=&\matr{F}_N\matr{B}_K\left(z+\sum_{ j \in\mathbb{Z}, j \neq0}\left(\sqrt{\gamma_{ j }}\matr{H}_{ j }\matrgr{\chi}_{ j }\right)\right)\\
	\equiv&\tran{\begin{bmatrix}
		z_1 & z_2 & \dots & z_N 
		\end{bmatrix}},
\end{align}
of which components are assumed to be $z_i~\mathcal{CN}\left(0,\sigma_{\matr{\tilde{z}}_i}^2 \right),\,\forall i \in \mathbb{N}_{\leq N}$ where $\sigma_{\matr{\tilde{z}}_i}^2$ is the noise and \ac{aci} power\footnote{\ac{isi} due to the previous \ac{ofdm} symbol transmitted by the desired user may also exist, however it is omitted as the system is modeled for a single \ac{ofdm} symbol for the sake of clarity. Interested readers may see the detailed multi-symbol system models provided in \cite{pekoz2017adaptive,Zhang2018}.} affecting $ i $th subcarrier that can be calculated per \cite{pekoz2017adaptive,Zhang2018}; \begin{equation}
	\sigma_{\matr{\tilde{z}}}^2\in \mathbb{R}^{+}_{N\times 1}=\tran{\begin{bmatrix}
	\sigma_{\matr{\tilde{z}}_1}^2 & \sigma_{\matr{\tilde{z}}_2}^2 & \dots & \sigma_{\matr{\tilde{z}}_N}^2
	\end{bmatrix}}
\end{equation} is the disturbance variance vector, and $\matrgr{\Theta}\in\mathbb{C}^{N\times N}$ is the complete \ac{cfr} matrix of the desired user's channel obtained as
 \begin{equation}
 \matrgr{\Theta}=\matr{F}_N\matr{B}_K\matr{H}_0\matr{A}\herm{\matr{F}_N}.\label{eq:cfrmat}
 \end{equation} 
 
  The diagonal components of \prettyref{eq:cfrmat} are the channel coefficients scaling the subcarrier in interest, and is referred to as the \ac{cfr} in the literature, hereinafter shown with $ \matrgr{\theta}\in \mathbb{C}^{N\times 1} $ where 
  \begin{equation}
  	\matrgr{\theta}=\diag{\matrgr{\Theta}},\label{eq:cfrvec}
  \end{equation}
  whereas the off-diagonal component on the $ k $th column of $ n\neq k $th row, would be the coefficient scaling the interference from the $ k $th subcarrier to the $ n $th subcarrier.
 Had there been no time variation in the channel and the \ac{med} of the channel was shorter than the discarded \ac{cp} duration at all times, that is,
 \begin{align}
 	h_{0,n_1,n_1-\Delta k}&=h_{0,n_2,n_2-\Delta k},\,&\forall n_1,n_2,\Delta k,\label{eq:invar}\\
 	h_{0,n,k}&=0,\,&\forall k<n-\left(L-K\right),\,\forall n,\label{eq:med}
 \end{align}
$\matr{B}_K\matr{H}_0\matr{A}$ would have resulted in a Toeplitz matrix, meaning
$ \matrgr{\Theta}$ would be a diagonal matrix, and the system would be \ac{ici} and \ac{isi}-free. Most modern receivers assume these conditions are valid and ignore \ac{ici} and \ac{isi}, which can only be estimated using advanced time-domain channel estimation algorithms such as \cite{kalakech_time-domain_2015}. 
Although results are numerically verified using signals received over time-varying vehicular channels, all algorithms, proposed or presented for comparison in this work, estimate the channel assuming \prettyref{eq:invar} and \prettyref{eq:med} are valid.
Equation \prettyref{eq:cfrvec} can also be written as 
\begin{equation}
\matrgr{\theta}=\matr{F}_{N}\matr{h},\,\forall n \in \mathbb{N}^{*}_{\leq N},\label{eq:cirtocfr}
\end{equation}
where the vector $ \matr{h}\in\mathbb{C}^{N\times 1}\triangleq \tran{\begin{bmatrix}\hat{h}_{0} & \hat{h}_{1} & \dots & \hat{h}_{N-1}\end{bmatrix}} $ is the static \ac{cir} estimate of the desired user's channel during that \ac{ofdm} symbol, of which elements in fact correspond to 
\begin{equation}
	\hat{h}_{k}=\frac{1}{N}\sum_{n=1}^{N} \matr{e}_{n,N} \matr{B}_K \matr{H}_0 \matr{A} \tran{\matr{e}_{\alpha,N}},
\end{equation}
where $\alpha = \left( \left( n-k-1 \right)\bmod N\right)+1$. 
Thus, ignoring the noise and \ac{aci} for the time being, if a known \ac{sc} symbol sequence denoted by $\matr{\tilde{d}}$, commonly referred to as the pilot sequence, is transmitted, the following symbols are expected to be received under aforementioned assumptions:
\begin{align}
	\matr{r}&=\sqrt{\gamma_{ 0 }}\tran{\matr{M}}\matr{F}_N\matr{B}_K\matr{H}_{ 0 }\matr{A}\herm{\matr{F}_{N}}\matr{M}\matr{\tilde{d}}\\
	&=\sqrt{\gamma_{ 0 }}\diag{\tran{\matr{M}}\matr{F}_{N}\matr{h}}\matr{\tilde{d}} \label{eq:recpilotshort} \\
	&=\sqrt{\gamma_{ 0 }}\diag{\matr{\tilde{d}}}\tran{\matr{M}}\matr{F}_{N}\matr{h}.\label{eq:recpilotleaveciralone}
\end{align}
Equation \prettyref{eq:recpilotleaveciralone} is an algebraic manipulation of \prettyref{eq:recpilotshort} in an effort to take the \ac{cir} outside the diagonalization for estimation in the next step. Assuming the receiver does not assume apriori knowledge of the \ac{snr} component and it is inherited within the \ac{cir}, the \ac{cir} estimate is obtained as
\begin{equation}
	\matr{h}=\left( \diag{\matr{\tilde{d}}}\tran{\matr{M}}\matr{F}_N \right)^{-1}\matr{r},\label{eq:orthcirest}
\end{equation}
wherein the inversion refers to the Moore-Penrose pseudoinverse.
The \acp{cir} of data carrying \ac{ofdm} symbols between pilot carrying \ac{ofdm} symbols are interpolated and according \ac{cfr} responses are calculated. Finally, equalized data symbol estimates are obtained as \cite{lucky1965automatic} 
\begin{equation}
	\matr{\hat{d}}=\tran{\matr{M}}\left(\left(\diag{\matrgr{\theta}\odot\conj{\matrgr{\theta}}+\sigma_{\matr{\tilde{z}}}^2}\right)^{-1}\conj{\diag{\matrgr{\theta}}}\matr{F}_N\matr{B}_{K}\matr{y}\right)
	\label{eq:zfeq}.
\end{equation}

\section{Proposed Method\label{sec:Proposed-Method}}
The Hann window must consist exactly of $N$ samples so that the spectrum is sampled at the right points as seen in \prettyref{fig:psd}. Furthermore, discarding the $L$ \ac{cp} samples at the beginning helps prevent \ac{isi} across consecutive desired \ac{ofdm} symbols transmitted by the desired user. 
The sample indices for the remaining samples can be written as $ \matr{n} =  \begin{bmatrix}
0 & 1 & \dots & N-1
\end{bmatrix} $
. The Hann window function normalized to window this interval without changing it's energy is obtained as
\begin{equation} \label{eq:hannfunc}
\matr{w}=\frac{4N}{2N+\sin{\left(\frac{\pi-2\pi N}{N-1}\right)}\csc{\left(\frac{\pi}{N-1}\right)}-1}\sin^{2}{\frac{\pi\matr{n}}{N-1}}.
\end{equation}
The Hann windowing matrix
 $ \matr{W}\in\mathbb{R}^{N\times \left(N+L\right)} $, that removes the \ac{cp} and windows the remaining received samples with the Hann function is formed as 
 \begin{equation} \label{eq:hannmat}
	\matr{W}=\begin{bmatrix}\matr{0}_{N\times L} & \diag{\matr{w}}\end{bmatrix}.
 \end{equation}
The received subcarrier vector $\matr{\tilde{r}}\in\mathbb{C}^{N\times1}$ that contains all Hann windowed subcarriers
is obtained as 
\begin{equation}
	\matr{\tilde{r}}=\matr{F}_N \matr{Wy}.\label{eq:hannwinrec}
\end{equation}

\subsection{\ac{ici} \& Channel Estimation in Hann \ac{rw-ofdm}}
A quick investigation of \prettyref{eq:hannfunc} and \prettyref{eq:hannmat} show that the orthogonality conditions presented in \cite{bala_shaping_2013} are not satisfied. In this subsection, we calculate the consequent \ac{ici} induced by the Hann window function, and accordingly engineer a method to estimate the \ac{cfr} and the disturbance variances using any pilot structure, including those of 4G \& 5G mobile communication.

Straightforward calculation reveals that 
\begin{align}\label{eq:hannComb}
	\matr{F}_N\diag{\matr{w}}\herm{\matr{F}_N}=\toep{\tran{\matrgr{\nu}}}{\matrgr{\nu}},
\end{align}
where $\matrgr{\nu}=\begin{bmatrix}
1 & -\sfrac{1}{2} & \matr{0}_{1\times \left(N-2\right)}
\end{bmatrix}$. 
Hence, assuming that \prettyref{eq:invar} and \prettyref{eq:med} are valid, the \ac{cfr} of the desired user's channel, if the Hann window is used, is given as
\begin{align}
	\matrgr{\tilde{\Theta}}&=\matr{F}_N\matr{W}\matr{H}_0\matr{A}\herm{\matr{F}_N}\\
	&=\toep{\tran{\matrgr{\nu}}}{\matrgr{\nu}}\matrgr{\Theta}.
\end{align}
Before the relevant subcarriers are demapped, note that Hann windowing causes received subcarriers that are adjacent to the edgemost pilot-transmitted subcarriers to carry copies of the pilots transmitted at these subcarriers. In an attempt to utilize this energy, this receiver demaps these subcarriers as well using an extended demapping matrix $ \tran{\check{\matr{M}}}\in\mathbb{Z}^{\left(D+2\right)\times N} $. Thus, ignoring channel disruption for the time being, if pilot symbols were transmitted, the received pilot symbols are obtained as \begin{align}
	\matrgr{\varDelta}&=\tran{\check{\matr{M}}}\matrgr{\tilde{\Theta}}\matr{M}\matr{\tilde{d}}\\
	&=\tran{\check{\matr{M}}}\toep{\tran{\matrgr{\nu}}}{\matrgr{\nu}}\matrgr{\Theta}\matr{M}\matr{\tilde{d}}\\
	&=\tran{\check{\matr{M}}}\toep{\tran{\matrgr{\nu}}}{\matrgr{\nu}}\diag{\matr{F}_{N}\matr{h}}\matr{M}\matr{\tilde{d}}\\
	&=\tran{\check{\matr{M}}}\toep{\tran{\matrgr{\nu}}}{\matrgr{\nu}}\diag{\matr{M}\matr{\tilde{d}}}\matr{F}_{N}\matr{h}.
\end{align}
The result of $\toep{\tran{\matrgr{\nu}}}{\matrgr{\nu}}\diag{\matr{M}\matr{\tilde{d}}}$, which can be thought as a filtering operation as it involves multiplication of pilot vector with a Toeplitz matrix as denoted in \prettyref{fig:blockdiag}, may include nulled pilots in some subcarriers due to induced \ac{ici} in case QPSK modulated Gold sequences are utilized as pilot signals \cite{3gpp.38.211}. However, the \ac{aci} rejection allows estimating the \ac{cir} better, namely, the disturbance level in
\begin{equation}
	\matr{h}=\left(\tran{\check{\matr{M}}}\toep{\tran{\matrgr{\nu}}}{\matrgr{\nu}}\diag{\matr{M}\matr{\tilde{d}}}\matr{F}_{N}\right)^{-1}\matrgr{\varDelta}\label{eq:hanncirest}
\end{equation}
is less than that of \prettyref{eq:orthcirest} if the \ac{aci} significantly outpowers \ac{awgn}. It is noteworthy that although  \prettyref{eq:hanncirest} still does not have a full-rank solution, exploiting the low density of $ \matr{h} $ as pointed out in \cite{Beek1995} allows implementation of an approximate \ac{lmmse} estimator relating the two sides as presented in \cite{Huang2010}. Any other variation of \cite{Beek1995} may be used, but \cite{Huang2010} is chosen in the numerical verification of this work since the computational complexity, error bounds and introduced delays of this approach remain within vehicular communication requirements as accepted by the community. Furthermore, depending on the ratio of nonzero pilot products to the delay spread, the estimation error can be shown to converge to zero\cite{Stewart1989}. The solution was later modified to be stable regardless of the condition of the pilot product matrix \cite{Vavasis1994} and also computationally highly efficient \cite{Hough1997}. Furthermore, the \ac{dft} of the disruption-only taps described in \cite{Huang2010} is used to estimate $ \matrgr{\sigma}_{\matr{\check{z}}}^{2} \in \mathbb{C}^{N\times 1}$, which is then interpolated throughout the data carriers similar to \ac{cir} estimates. After the \ac{cir} estimates are interpolated, they are transformed to frequency domain to obtain the \ac{cfr} estimates $ \hat{\matrgr{\theta}} \in \mathbb{C}^{D\times 1}$.

\subsection{Design of an \ac{mrc}-\ac{sic} Receiver}
Similar to the described channel estimation, this receiver also attempts to utilize the energy in the subcarriers adjacent to the edgemost data carriers. In this case, the received symbols $ \check{\matr{d}}\in\mathbb{C}^{\left(D+2\right)\times1} $ are written as 
\begin{align}
	\check{\matr{d}}&=\tran{\check{\matr{M}}}\matr{r}\\
&=\matr{\tilde{H}}\matr{d}+\tran{\check{\matr{M}}}\matr{F}_N\matr{W}{\matr{z}},
\end{align}
where, the extended effective
channel matrix $\matr{\tilde{H}}\in\mathbb{C}^{\left(D+2\right)\times D}$
is obtained as
\begin{dmath} \label{eq:extchan}
 \matr{\tilde{H}}=\toep{\tran{\begin{bmatrix}\sfrac{-1}{2} & \tilde{\matrgr{\nu}} \end{bmatrix}}}{\begin{bmatrix}\sfrac{-1}{2} & \matr{0}_{1\times\left(D-1\right)}\end{bmatrix}}\diag{\hat{\matrgr{\theta}}},
\end{dmath}
where $\tilde{\matrgr{\nu}}=\begin{bmatrix}
	1 & -\sfrac{1}{2} & \matr{0}_{1\times \left(D-1\right)}
\end{bmatrix}$.
The energy due to the signal modulated to the $m$th transmitted subcarrier on the $k$th observed subcarrier is in the $k$th row and $m$th column of $\matrgr{\Sigma}\in\mathbb{R}^{\left(D+2\right)\times D}$ where
\begin{equation}
	\matrgr{\Sigma}= \osq{\matr{\tilde{H}}}.
\end{equation}
The signal-plus-\ac{ici} power on the $k$th observed subcarrier is given in the $k$th column
 of $\matrgr{\sigma}\in\mathbb{R}^{1\times \left(D+2\right)}$, where
  \begin{equation}
 \matrgr{\sigma}=\matr{1}^{1\times D}\tran{\left(\matrgr{\Sigma}\right)}. 
 \end{equation}
If the $m$th transmitted subcarrier is in interest, the disruption-plus-\ac{ici} power contribution that would come from combining the $k$th observed subcarrier with unit gain is given on the $m$th row and $k$th column of 
\begin{equation}\label{eq:naimat}
 \hat{\matrgr{\Sigma}}=\left(\kron{\matr{1}^{D\times1}}{\left(\tran{\matrgr{\sigma}_{\check{\matr{z}}}^{2}}+\matrgr{\sigma}\right)}\right)-\tran{\left(\matrgr{\Sigma}\right)},
\end{equation}
where $\hat{\matrgr{\Sigma}}\in\mathbb{R}^{D\times\left(D+2\right)}$ and $ \matrgr{\sigma}_{\check{\matr{z}}}^{2} \in \mathbb{R}^{D+2\times 1} $ is the disruption variance vector at the output of the extended demapper. The \ac{mrc} matrix is then\cite{kahn1954ratiosquarer}\begin{equation}
  \matr{\tilde{C}}=\herm{\matr{\tilde{H}}}\odot\left(\tran{\matrgr{\Sigma}}\oslash\hat{\matrgr{\Sigma}}\right),
  \end{equation}
where $\matr{\tilde{C}}\in\mathbb{C}^{D\times \left(D+2\right)}$.
Although $\tilde{\matr{C}}$ maximizes the \ac{sinr}, the resulting data estimates $\tilde{\matr{C}}\matr{\check{d}} $ would be scaled with non-unity coefficients.
The "equalized" \ac{mrc} matrix $\matr{C}\in\mathbb{C}^{D\times \left(D+2\right)}$ is obtained as 
\begin{equation}
 \matr{C}=\matr{\tilde{C}} \oslash \left(\kron{\matr{1}^{1\times \left(D+2\right)}}{\diag{\matr{\tilde{C}}\matr{\tilde{H}}}}\right).
 \end{equation}
The symbol estimates at the \ac{mrc} output $\matr{\breve{d}}\in\mathbb{C}^{D\times1} $ is
 \begin{equation}
	\matr{\breve{d}}=\matr{C}\matr{\check{d}}.\label{eq:mrcout}
\end{equation}
The post-\ac{mrc} gain of the \ac{ici} component present on the $m$th subcarrier due to the
$k$th subcarrier is given on the $m$th row and $k$th
column of  $ \matr{G}\in\mathbb{C}^{D\times D} $ where
\begin{equation} \matr{G}=\matr{C}\matr{\tilde{H}}-\matr{I}^{D}. \label{eq:postmrcicicoeff}\end{equation}
The disruption power accumulated on the $m$th
subcarrier after \ac{mrc} is given on the $m$th column of 
\begin{equation}
\matrgr{\rho}=\lvert \matr{C} \rvert^{2}\matrgr{\sigma}_{\check{\matr{z}}}^{2}, \label{eq:mrcnoiseout}
\end{equation}
where  $ \matrgr{\rho}\in\mathbb{R}^{1\times D} $. $\matr{\breve{d}}$, $\matr{G}$ \& $\matrgr{\rho} $ are fed
to the SISO decoder in \cite[Sec V]{barbieri2009timefrequency}, and
the soft decision turbo \ac{sic} equalizer described thereon is used
to obtain symbol estimates $\matr{\hat{d}}$.

\subsubsection{Computational Complexity}
The derivation of the \ac{mrc} operation may create an impression that it requires series of sequential operations. Although these steps are detailed for the derivation, the implementation complexity is limited as a result of the limited number of interference terms. For example, if the index of the extended demapped subcarriers are considered to be $0$ and $D+1$ for the sake of brevity in this context, the $d\in\mathbb{Z}_{1<d<D}$th term of \prettyref{eq:mrcout} is explicitly stated as
\begin{equation}
	\breve{\matr{d}}_{d}=\frac{\sum_{ \kappa=d-1 }^{d+1} \tilde{\matrgr{\gamma}}_{d,\kappa}\conj{\matr{\tilde{H}}_{d,\kappa}} \matr{\check{d}}_{\kappa} }{\sum_{ \kappa=d-1 }^{d+1} \tilde{\matrgr{\gamma}}_{d,\kappa}\lvert \matr{\tilde{H}}_{d,\kappa} \rvert^{2} }, \label{eq:mrcoutexplicit}
\end{equation}
where \begin{equation}
	\tilde{\matrgr{\gamma}}_{d,\kappa}= \frac{ \lvert \matr{\tilde{H}}_{d,\kappa} \rvert^{2} }{ \matrgr{\sigma}_{\matr{\check{z}}_{\kappa}}^{2} + \sum_{ \substack{\tau\in\left\{\kappa-1,\kappa,\kappa+1\right\} \\ \tau\neq d}} \lvert \matr{\tilde{H}}_{\tau,\kappa} \rvert^{2} } \label{eq:hannmrcsinr}
\end{equation}
is the \ac{sinr} of the symbol transmitted in the $d$th subcarrier at the $\kappa$th received subcarrier. Noting that the off-diagonal components of $ \tilde{\matr{H}} $ can be obtained from $ \matrgr{\theta}$ using simple bit operations, calculation of $ \lvert \matr{\tilde{H}} \rvert^{2} $ is ignored as well as the channel estimation using \cite{Huang2010,Hough1997} and  Fourier transform in \prettyref{eq:hannwinrec} since they are included in all algorithms. The complexity of the rest of the steps are provided in \prettyref{tab:compcomplex}, where $M$ is the cardinality of the used constellation and the approximations refer to the cases where constant magnitude (\ac{psk}) constellations are used.
\begin{table}
	\caption{Computational Complexity of Algorithm Steps}\label{tab:compcomplex}
	\centering
	\begin{tabular}{ c | c | c }
		\hline
		\bfseries Step & \bfseries Real~Multiplications & \bfseries Real~Additions \\
		\hline\hline
		\prettyref{eq:hannwinrec} & $2N$ & $0$ \\
		\hline
		\prettyref{eq:hannmrcsinr} & $3D$ & $6D$ \\
		\hline
		\prettyref{eq:mrcout} & $22D$ & $12D$ \\
		\hline
		\prettyref{eq:postmrcicicoeff} & $24D$ & $16D$ \\
		\hline
		\prettyref{eq:mrcnoiseout} & $9D$ & $5D$ \\
		\hline
		A priori probabilities & $ 4MD $ & $ 3MD $ \\
		\hline\hline
		\bfseries Equalized~MRC~Total & $2N+(58+4M)D$ & $(39+3M)D$ \\
		\hline\hline
		\cite[Sec. V, $\mu $]{barbieri2009timefrequency} & $3MD$ & $2(M-1)D$ \\
		\hline
		\cite[Sec. V, $\sigma^{2} $]{barbieri2009timefrequency} & $(M+2)D \approx 2D $ & $(M+1)D$ \\
		\hline
		\cite[Sec. V, $\tilde{y} $]{barbieri2009timefrequency} & $16D $ & $16D$ \\
		\hline
		\cite[Sec. V, $\Sigma $]{barbieri2009timefrequency} & $4D $ & $4D$ \\
		\hline
		Extrinsic probabilities & $ 4MD $ & $ 3MD $ \\
		\hline\hline
		\bfseries Each~SIC~Iteration & \begin{tabular}{r}
			$(22+4M)D$\\ $\approx (22+3M)D$
		\end{tabular} & $(19+3M)D$ \\
		\hline
	\end{tabular}
\end{table}

Note that the proposed method does not include any non-linear or sequential operation, hence it is possible to obtain the extrinsic probabilities at the end of any number of \ac{sic} iterations in a single clock cycle if the memory and hardware architecture allows \cite[Sec. V]{barbieri2009timefrequency}.

\section{Numerical Verification\label{sec:Numerical-Results}}

The gains of Hann windowing \ac{ofdm} receivers are shown using numerical simulations and compared to other methods. The assumptions advised in \cite{3gpp.38.802} for the 3GPP \ac{nr} band "n41" \cite{3gpp.38.101-1} and a system bandwidth of \SI{50}{\mega\hertz} were used. 
There are two identical interfering users each utilizing the bands on either side of the band occupied by the desired user. Both interfering users' experience channels with \SI{20}{\deci\bel} SNR having \ac{tdl}-C \ac{pdp} with \SI{300}{\nano\second} RMS delay spread and mobility \SI[per-mode=symbol]{3}{\kilo\meter\per\hour}.
The desired user's channel has the \ac{tdl}-A \ac{pdp} with \SIlist{10;30}{\nano\second} RMS delay spread \cite{3gpp.38.901}, mobility \SI[per-mode=symbol]{120}{\kilo\meter\per\hour} and was evaluated for \SIrange{10}{30}{\deci\bel} \ac{snr}. The guard bands between users also vary from \SIrange{30}{105}{\kilo\hertz}. The desired user has a subcarrier spacing of \SI{60}{\kilo\hertz} corresponding to $N=1024$, whereas both interfering users have subcarrier spacings of \SI{15}{\kilo\hertz} corresponding to 4096-\ac{fft}. Starting from the fourth symbol, all subcarriers of every seventh OFDM symbol of the desired user is loaded with \ac{pusch} \ac{dmrs} symbols defined in \cite{3gpp.38.211}. The desired user utilizes $D=12$ subcarriers in the remaining symbols to convey data using the same \ac{mcs} for all SNR values which consists of QPSK modulation and $\left(51/63\right)\times\left(7/16\right)$ standard \cite{Bose1960, Hocquenghem1959} and extended \cite{peterson1965onthe} \ac{bch} \ac{tpc} \cite{pyndiah1998nearoptimum}. The interfering users utilize 1632 subcarriers each throughout the whole communication duration. There is also 128 samples time offset between the the desired user and both interfering users. The bit probabilities are calculated using approximate \acp{llr} for all receivers.

Both interfering users transmit $2$-continuous \ac{ofdm} for the results labeled with NC-OFDM, while the desired user transmits $1$-continuous \ac{ofdm} and the receiver performs 6 iterations to estimate the transmitted correction vector \cite{beek2009ncontinuous} and cancel it. Both interfering users perform transmit filtering in results labeled with \ac{fofdm} as described in \cite{HuaweiFOFDM2016} while the samples of desired user are match filtered, where tone offset values are set to the respective guard band of that simulation for all users. For all other results, both interfering users utilize normal \ac{cp} overhead and window the \ac{isi}-free \ac{cp} samples at the transmitter using \ac{ssw} functions optimized to maximize their frequency localization\cite{guvenkaya2015awindowing}.
The desired user employs normal \ac{cp} overhead for all cases and the \ac{isi}-free samples are utilized for \ac{ssw} maximizing \ac{aci} rejection \cite{guvenkaya2015awindowing} in the results labeled as \ac{rw-ofdm}. For the results labeled as adaptive \ac{rw-ofdm} (ARW-OFDM), the window duration of each subcarrier is determined per \cite{pekoz2017adaptive}. A total of 6 \ac{sic} iterations are performed for the Hann windowing receivers and the \ac{ber} values at the output of each iteration are obtained and presented in the \ac{ber} results. The number of iterations are denoted accordingly, and not all iterations were presented in all \ac{ber} results for the sake of clarity. Furthermore, the theoretical \ac{ber} bound achievable by an Hann windowing receiver if \ac{ici} is cancelled perfectly is obtained from the Hann-windowed channel disruption and presented with the label Hann-Theory. It should be noted that the receiver design featured in this work is suboptimal in most cases, and is only presented to demonstrate the concept using a receiver that is suitable for the low-latency requirements of \ac{urllc}. Non-linear or decision directed receivers that  consistently achieve theoretical \ac{ber} bound are left as future work.

\begin{figure}
	\subfloat[]{
	\includegraphics{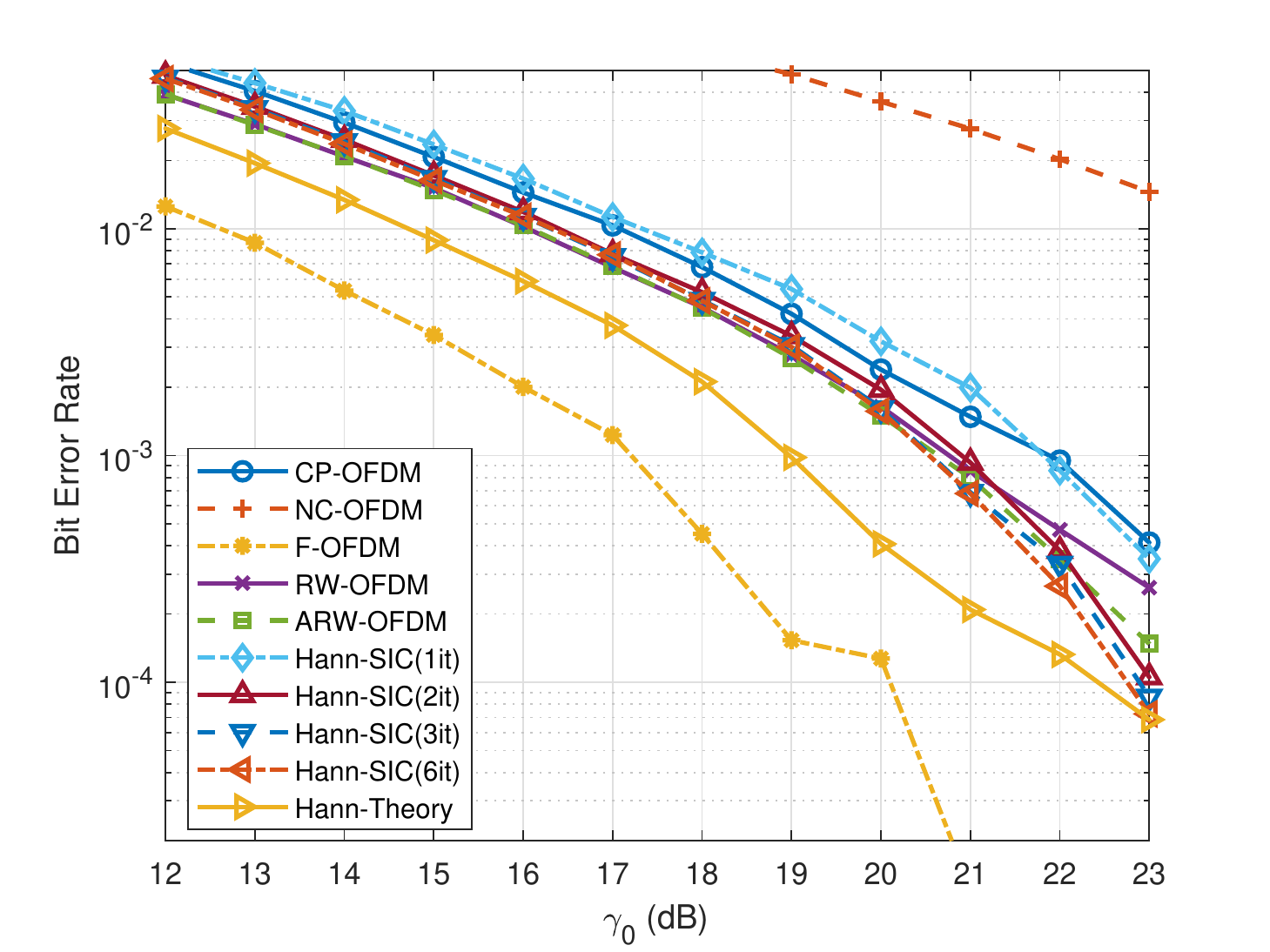}
	\label{fig:2GB10ns}}\hfil
\subfloat[]{
\includegraphics{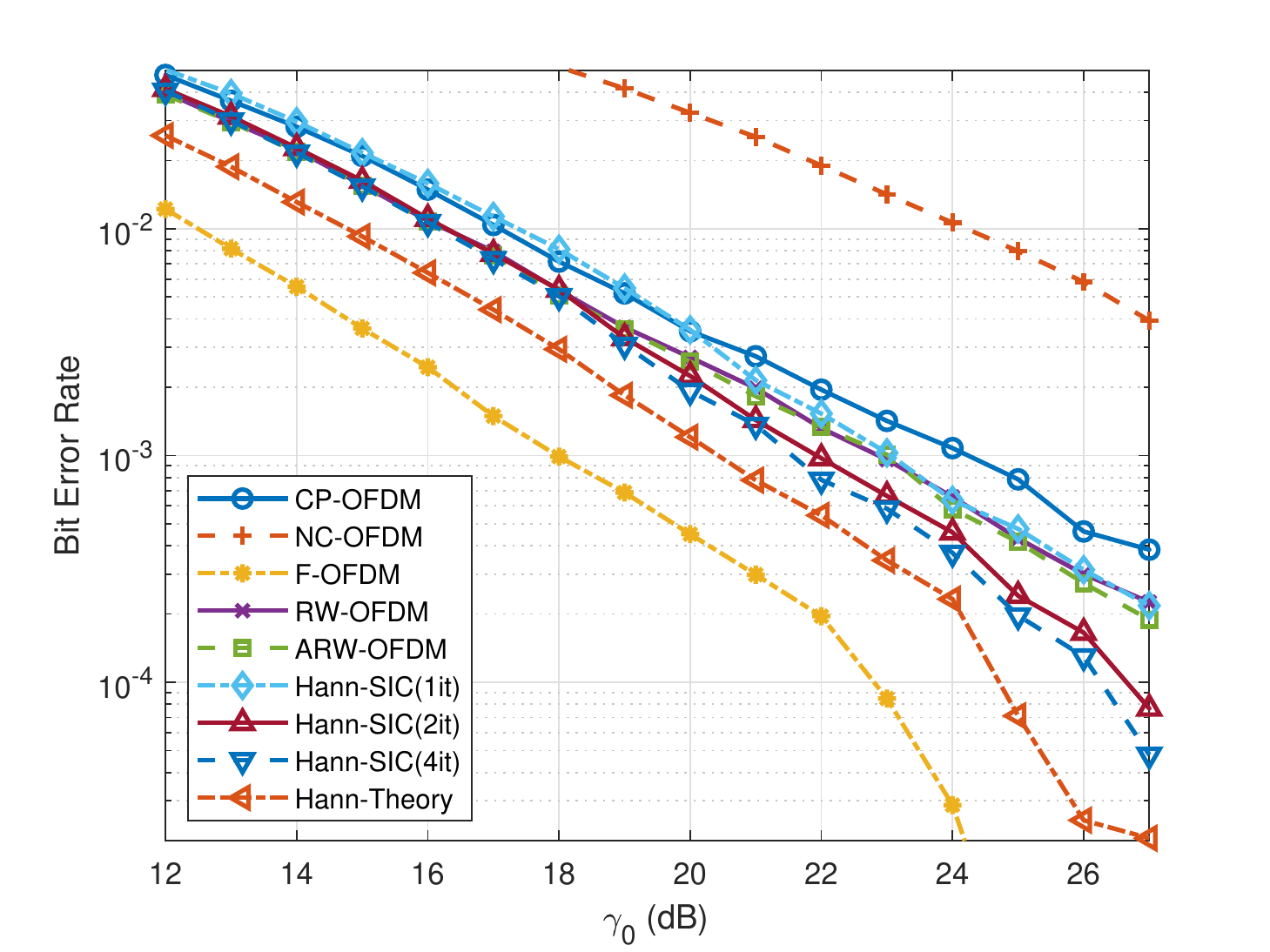}
\label{fig:2GB30ns}}
\caption{\acp{ber} of various transceivers for guard bands of \SI{30}{\kilo\hertz} between each user and (a) \SI{10}{\nano\second} and (b) \SI{30}{\nano\second} RMS delay spread.}\label{fig:bercurves2}
\end{figure}

\begin{figure}
\subfloat[]{
\includegraphics{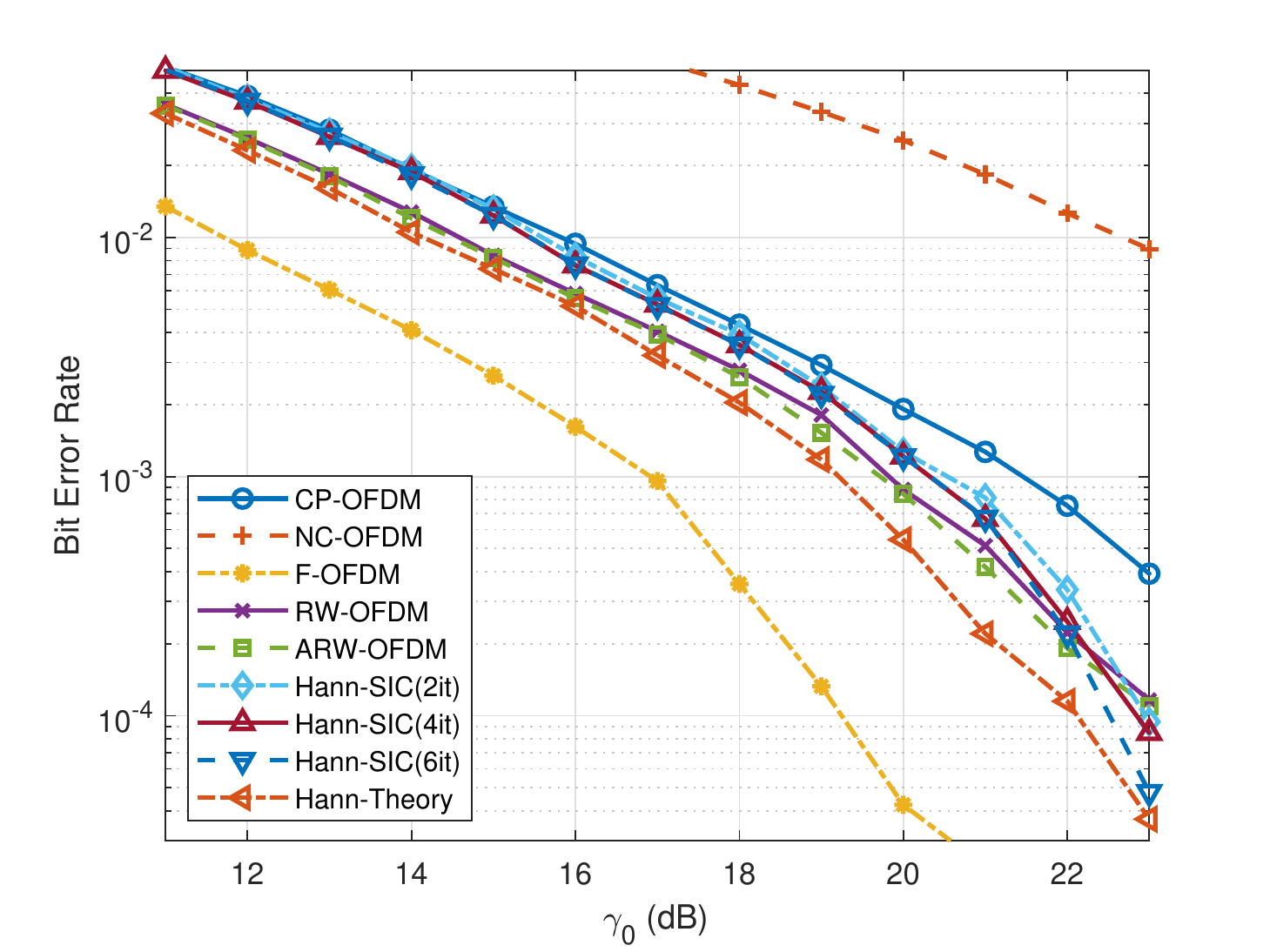}
\label{fig:7GB10ns}}\hfil
\subfloat[]{
\includegraphics{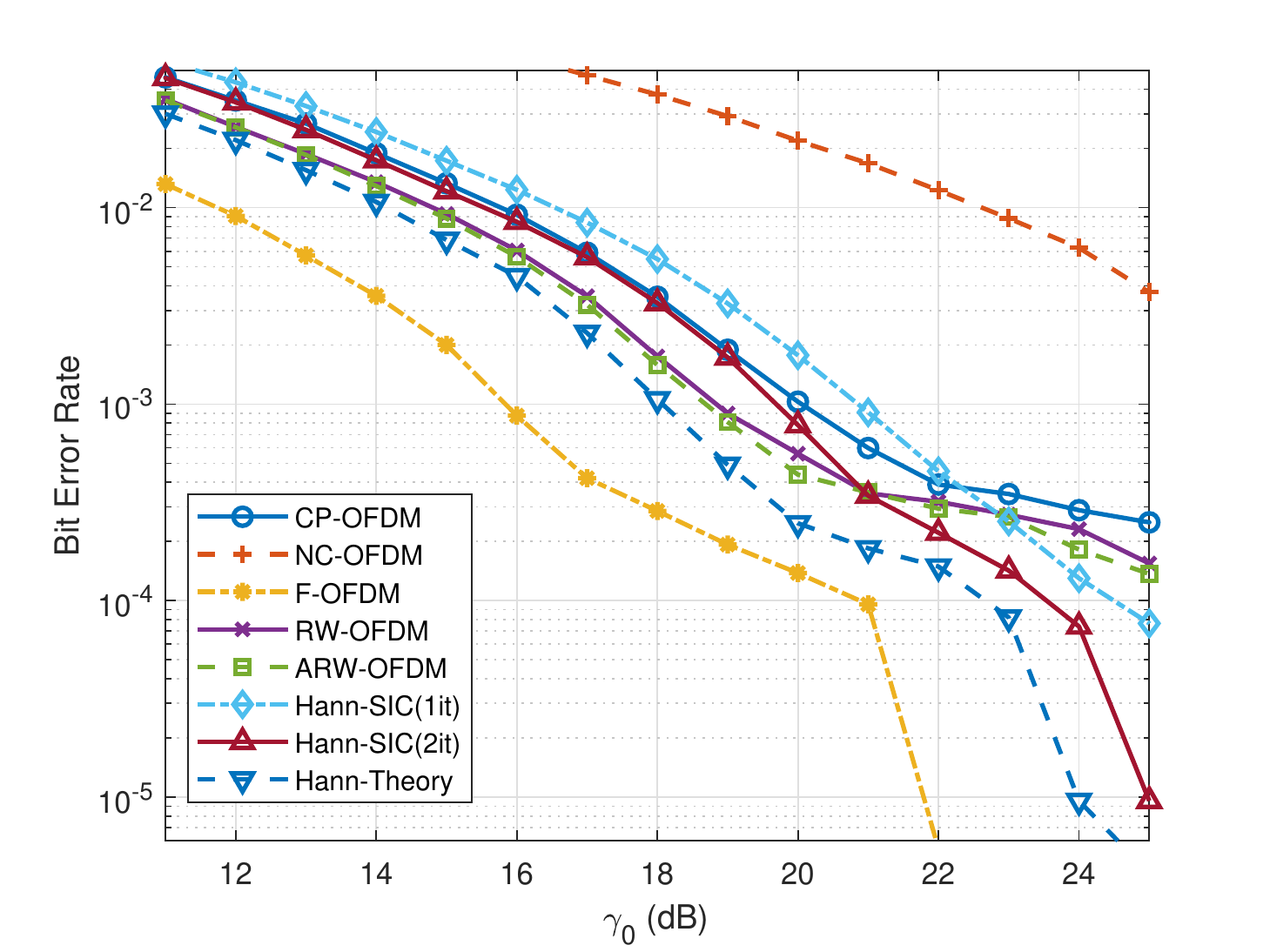}
\label{fig:7GB30ns}}
\caption{\acp{ber} of various transceivers for guard bands of \SI{105}{\kilo\hertz} between each user and (a) \SI{10}{\nano\second} and (b) \SI{30}{\nano\second} RMS delay spread.}\label{fig:bercurves7}
\end{figure}

\prettyref{fig:2GB10ns} demonstrates that for little guard band and short delay spread, orthogonal windowing algorithms outperform Hanning receivers for the low SNR regions, as low SINR prevents successful ICI estimation and cancellation. However, Hanning receivers with as little as 3 iterations achieve the target $10^{-3}$ BER earlier than orthogonal windowing algorithms and experience the so-called BER waterfall at a lower SNR threshold than compared to orthogonal windowing algorithms. As SNR increases further, even 2 iterations are sufficient to outperform orthogonal windowing algorithms while as little as 6 iterations allow rates very close to the theoretical limit. The motivation behind Hanning receivers become clearer in \prettyref{fig:2GB30ns} as delay spread elongates. As orthogonal windowing algorithms lose the advantage of longer windowing durations, their rates shift closer to the baseline rectangular receiver, while the effect on Hanning receivers remain limited to increased fading. Hanning receiver with only 1 iteration show the same performance as orthogonal windowing receivers performance for the high SNR region, while as little as 2 iterations outperform the orthogonal windowing receivers at all SNR values. The only observable effect on Hanning receivers is the shift of the waterfall threshold to higher SNRs.
Increasing the guard band drastically reduces the ACI present on the desired user's band and narrows the gap between all algorithm as seen in \prettyref{fig:bercurves7}. In \prettyref{fig:7GB10ns}, compared to \prettyref{fig:2GB10ns}, the increase in the guard band extended the orthogonal windowing algorithm's lead against Hanning receivers beyond the target BER. However, it is seen that Hanning receivers, although requiring one more iteration, still outperform orthogonal windowing algorithms. Similarly as delay spreads elongate in \prettyref{fig:7GB30ns}, the Hanning receiver's advantage becomes more obvious with waterfall threshold moving further to lower SNRs compared to orthogonal windowing algorithms.

Noting that the ACI sources utilize either transmitter \ac{w-ofdm} or \ac{fofdm} in other scenarios, both having superior \ac{oob} emission suppression compared to \ac{ncofdm}, the need to properly estimate and cancel the correction vector limits the BER performance of \ac{ncofdm}. While \ac{fofdm} has better \ac{ber} performance beyond that theoretically achievable by Hanning receivers, Hanning receivers are used to resolve conventional CP-OFDM signals whereas \ac{fofdm} can only be used to receive signals transmitted using a \ac{fofdm} transmitter. Particularly, the filter lengths are $N/2+1$ per \cite{HuaweiFOFDM2016}, the computational complexity of \ac{fofdm} is $\left(N+L\right)\left(2N+4\right)$ real multiplications and $ \left(N+L\right)\left(3N/2+2\right)$ real additions at both transmitter and receiver accordingly as filters consist of complex values. Similarly, the \ac{ssw} \ac{rw-ofdm} scheme described in \cite{guvenkaya2015awindowing} requires $6KD$ real multiplications and $4KD$ real additions to estimate received symbols. It is noteworthy that the complexity of \ac{fofdm} scales on the order of FFT-size squared, whereas the complexity of windowing receivers scale linearly with the number of data-carrying subcarriers. Therefore, windowing receivers are particularly important for narrowband communications. The receiver computational complexity of Hanning receivers are accordingly compared with \ac{fofdm} and RW-OFDM in \prettyref{tab:complexcompare}. Note that \ac{fofdm} has symmetric computational complexity between the transmitting and receiving devices, whereas windowing receivers do not change the transmitter structure and do not cause any computational burden at the transmitting device. The window duration for RW-OFDM in short delay spread is considered as it makes more sense to apply RW-OFDM in that case.
\begin{table}
	\caption{Computational Complexity Comparison}\label{tab:complexcompare}
	\centering
	\begin{tabular}{c | c | c}
		\bfseries Algorithm & \bfseries Real Mult. & \bfseries Real Add. \\
		\hline\hline
		\ac{fofdm} & 2,248,992 & 1,685,648 \\
		\hline
		RW-OFDM & 3,672 & 2,448 \\
		\hline
		Hann-SIC(1it) & 3,224 & 984\\
		\hline
		Hann-SIC(2it) & 3,632 & 1,356\\
		\hline
		Hann-SIC(3it) & 4,040 & 1,728\\
		\hline
		Hann-SIC(4it) & 4,448 & 2,100\\
		\hline
		Hann-SIC(6it) & 5,264 & 2,844\\
		\hline
	\end{tabular}
\end{table}
\section{Conclusion\label{sec:Conclusion}}

ACI is a critical problem in 5G and beyond scenarios due to the coexistence of \ac{ofdm} based non-orthogonal signals. To tackle the \ac{aci} problem, we propose a novel Hann window function based low complexity receiver windowing method that is fully compatible with the frame structure of existing standards and needs no redundancy in the signal and no modifications on the transmitting devices. The proposed method improves the achievable capacity in the presence of high power non-orthogonal signals on adjacent channels when it is coupled with simple interference mitigation techniques. The proposed method allows superior ACI rejection and reducing guard bands without requiring extensions, and on the contrary, allows shortening the currently used extension for future higher mobility applications. Although the gap between prior art and the proposed methods widens with increasing delay spread and decreasing guard bands, the proposed methods outperform prior art in short delay spreads and large guard bands as well.
This study paves the way towards future standard compliant \ac{aci} rejection research by showing gains of a simple receiver, inspiring sophisticated algorithms that outperform the presented by achieving performance bounds with less receiver complexity.

\bibliographystyle{IEEEtran}

\begin{IEEEbiography}[\includegraphics{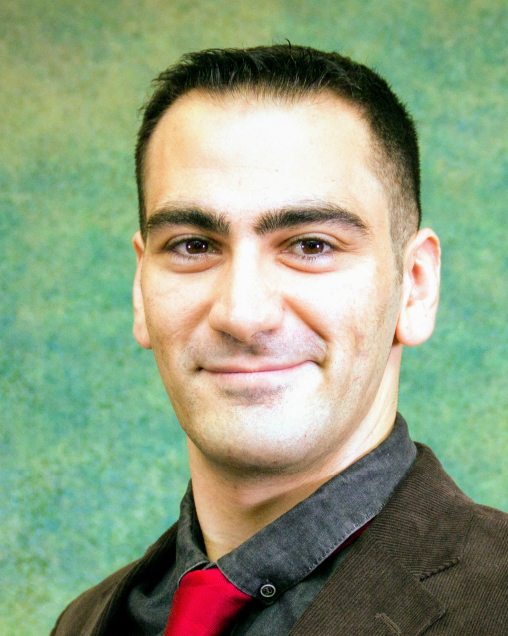}]{Berker Peköz}
(GS'15) received the B.S. degree (Hons.) in electrical and electronics engineering from \acl{metu}, Ankara, Turkey in 2015, and the M.S.E.E. from \acl{usf}, Tampa, FL, USA in 2017.
He is currently pursuing the Ph.D. degree at University of South Florida, Tampa, FL, USA. 
His research is concerned with backward compatible standard compliant waveform design, and PHY algorithms and optimization thereof.
Mr. Peköz is a member of Tau Beta Pi.
\end{IEEEbiography}

\begin{IEEEbiography}[{\includegraphics{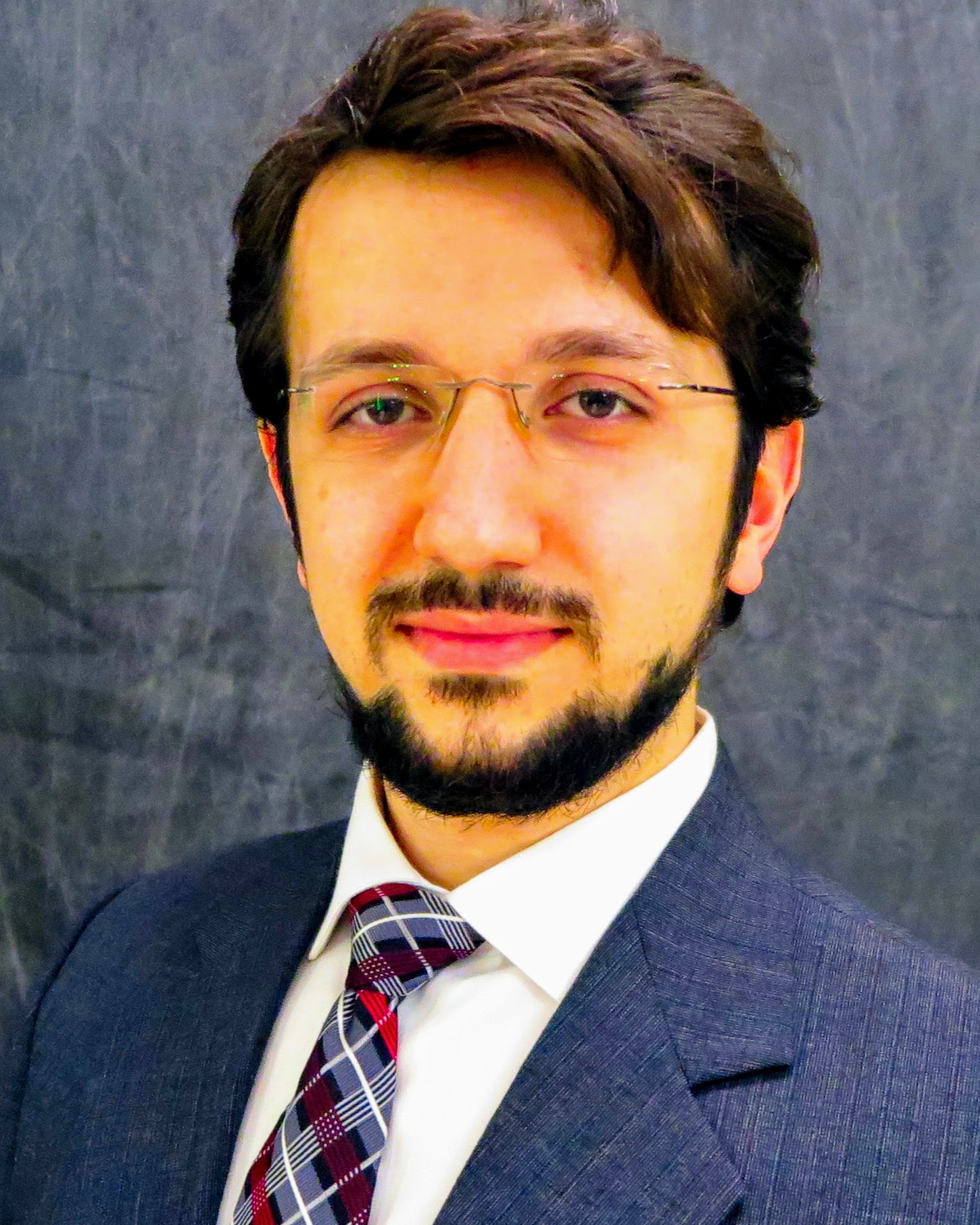}}]{Zekeriyya~Esat~Ankaralı} (GS'16) received the B.Sc. degree (Hons.) in control engineering from Istanbul Technical University, Istanbul, Turkey, in 2011, and the M.Sc. and Ph.D. degrees both in electrical engineering from the University of South Florida, Tampa, FL, USA, in 2013 and 2017, respectively. 
Since 2018, he has been with MaxLinear Inc., Carlsbad, CA, USA,  as a staff communications systems engineer
. 
His research interests are waveform design, 
physical layer security, and in vivo communications.
	
\end{IEEEbiography}

\begin{IEEEbiography}[\includegraphics{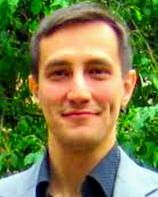}]{Selçuk Köse}
(S’10--M’12) received the B.S. degree in electrical and electronics
engineering from Bilkent University, Ankara, Turkey, in 2006, and
the M.S. and Ph.D. degrees in electrical engineering from the University
of Rochester, Rochester, NY, USA, in 2008 and 2012, respectively.
He was an Assistant Professor of Electrical Engineering at the University of South Florida, Tampa, FL, USA. He is currently an Associate Professor of Electrical and Computer Engineering at University of Rochester, Rochester, NY, USA.
Dr. Köse is an Associate Editor
of the \emph{World Scientific Journal of Circuits, Systems, and Computers}
and the \emph{Elsevier Microelectronics Journal}. 
\end{IEEEbiography}

\begin{IEEEbiography}[\includegraphics{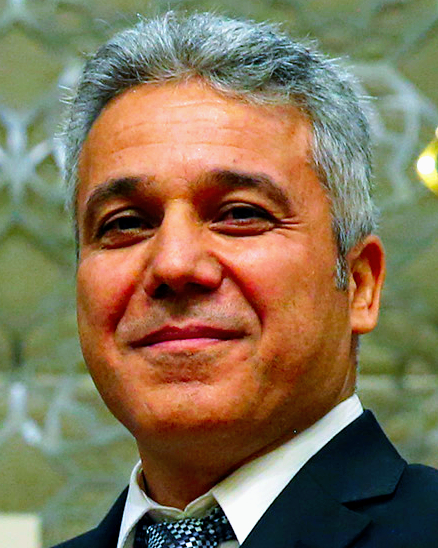}]{Hüseyin Arslan}
(S'95-M'98-SM'04-F'15) received the B.S. degree in electrical and
electronics engineering from \acl{metu}, Ankara, Turkey in 1992,
and the M.S. and Ph.D. degrees in electrical engineering from Southern
Methodist University, Dallas, TX, USA in 1994 and 1998, respectively.
From 1998 to 2002, he was with the research group of
Ericsson Inc., Charlotte, NC, USA. He is currently a Professor of Electrical Engineering at the University of South Florida, Tampa, FL, USA, and the Dean of the College of Engineering and Natural
Sciences at the İstanbul Medipol University, İstanbul, Turkey.
Dr. Arslan is currently a member of the editorial board
for the \textsc{IEEE Communications Surveys and Tutorials} 
and the \emph{Sensors Journal}. 
\end{IEEEbiography}

\end{document}